\documentclass{ws-procs975x65}            
\begin{document}
\title{Strongly interacting matter at RHIC:
experimental highlights}

\author{V. A. Okorokov}

\address{Physics Department, National Research Nuclear University ``MEPhI" \\
(Moscow Engineering Physics Institute),\\
Moscow, 115409, Russia\\
E-mail: VAOkorokov@mephi.ru;~Okorokov@bnl.gov}

\begin{abstract}
Recent experimental results obtained at the Relativistic Heavy-Ion
Collider (RHIC) will be discussed. Investigations of different
nucleus-nucleus collisions in recent years focus on two main
tasks, namely, the detailed study of sQGP properties and the
exploration of the QCD phase diagram. Results at top RHIC energy
provide important information about event shapes as well as
transport and thermodynamic properties of the hot medium for
various flavors. Heavy-ion collisions are a unique tool for the
study of topological properties of theory. Experimental results
obtained for discrete QCD symmetries at finite temperatures are
discussed. These results confirm indirectly the topologically
non-trivial structure of the QCD vacuum. Most results obtained
during phase-I of the RHIC beam energy scan (BES) program show
smooth behavior vs initial energy. However, certain results
suggest the transition in the domain of dominance of hadronic
degrees of freedom at center-of-mass energies between 10-20 GeV.
Future developments and more precise studies of features of the
QCD phase diagram in the framework of phase-II of RHIC BES will be
briefly discussed.
\end{abstract}


\bodymatter

\section{Introduction}\label{vao:sec1}
Investigation of strongly interacting matter under extreme
conditions is of value for various fields of fundamental physics,
namely for quantum chromodynamics (QCD), cosmology and
relativistic astrophysics. The Relativistic Heavy-Ion Collider
(RHIC) at Brookhaven National Laboratory was designed and was
built for investigations in the field of stong-interaction physics
specially. The data samples taken since 2000 are shown in Table
\ref{table:1}. The main result of the first stage of the heavy-ion
program at RHIC is the demonstration that new state of nuclear
matter -- strongly coupled quark-gluon plasma (sQGP) -- has been
observed in nucleus-nucleus collisions at top RHIC energy and
mid-rapidity. Subsequent experimental goals focus on studying the
properties of the sQGP and exploring features of the phase diagram
of strongly interacting matter.

\begin{table}
\tbl{Data samples obtained during RHIC runs.}
{\begin{tabular}{@{}cc@{}} \toprule Species &
$\sqrt{s_{\mathrm{NN}}}$, GeV\\\colrule
$\mbox{p+p}^{\text a}$     & 22.0$^{\text b}$, 62.4, 200, 410$^{\text b}$, 500, 510 \\
$\mbox{d+Au}$              & 200 \\
$\mbox{He$^{3}$+Au}$       & 200 \\
$\mbox{Cu+Cu}$             & 22.4$^{\text b}$, 62.4, 200 \\
$\mbox{Cu+Au}$             & 200 \\
$\mbox{Au+Au}$             & 7.7, 9.2$^{\text b}$, 11.5, 14.5, 19.6, 27.0, 39.0, 55.8$^{\text b}$, 62.4, 130, 200 \\
$\mbox{U+U}$               & 193 \\\botrule
\end{tabular}}
\begin{tabnote}
$^{\text a}$ with unpolarized ($\sqrt{s}=62.4$ GeV) and with
longitudinal / transverse polarized beams; $^{\text b}$ run with small integral luminosity\\
\end{tabnote}
\label{table:1}
\end{table}

\section{Probing of the sQGP at high RHIC energies}\label{vao:sec2}
During recent years the study of event shapes at higher RHIC
energies ($\sqrt{s_{\mathrm{NN}}} \geq 62.4$ GeV) is focused on
measurements of $v_{2}$ up to high
$p_{\,\mathrm{T}}$,\cite{PHENIX-PRC-80-024909-2009} higher-order
harmonics,\cite{PHENIX-PRL-105-062301-2010} comparisons of various
collisions, \cite{STAR-PRC-81-044902-2010} investigation of
scaling behavior of $v_{2}$ on number of constituent quarks
($n_{\mathrm{Q}}$) in the coalescence regime.
\cite{PHENIX-PRC-85-064914-2012} Latest results indicate that the
particle production is dominated by parton recombination in the
central collisions at top RHIC energy and intermediate
$p_{\,\mathrm{T}}$, but recombination is not the dominant mode in
noncentral collisions at $p_{\,\mathrm{T}} > 2$ GeV/$c$ and some
other mechanisms -- parton-energy loss, jet chemistry, and
different fragmentation functions -- may contribute to generating
the observed azimuthal anisotropy of particle emission. The
$v_{2,4}$ of $\pi^{0}$ and $\eta$ mesons with high
$p_{\,\mathrm{T}}$ exhibit the possibility of the combined
influence of initial-geometry fluctuations of collisions and
finite viscosity on space-time evolution of created matter.
\cite{PHENIX-PRC-88-064910-2013} Finite $v_{2}$ found for direct
$\gamma$ in the thermal region \cite{PHENIX-PRL-109-122302-2012}
supports the hypothesis of early thermalization and small
viscosity for hot matter created in central $\mbox{Au+Au}$
collisions at $\sqrt{s_{\mathrm{NN}}}=200$ GeV. The $v_{2}$ in
$\mbox{d+Au}$ at RHIC \cite{PHENIX-PRL-111-212301-2013} agrees
qualitatively both with results for $v_{2}$ in $\mbox{p+Pb}$ at
the LHC and with hydrodynamics.

Dihadron azimuthal correlations as a function of the trigger
particle's azimuthal angle relative to the event plane at top RHIC
energy with additional comparison of small ($\mbox{d+Au}$) and
large ($\mbox{Au+Au}$) systems \cite{STAR-PRC-89-041901-2014} show
the presence of path--length--dependent jet quenching. This
conclusion is confirmed by both the results for $\pi^{0}$
azimuthal anisotropy and the azimuthal correlations in
$h^{\pm}-\pi^{0}$ pairs as a function of the neutral trigger's
orientation with respect to the reaction plane.
\cite{PHENIX-PRL-105-142301-2010} Study of the three-particle
azimuthal correlations for various systems at top RHIC energy
signals conical emission of charged hadrons correlated with high
$p_{\,\mathrm{T}}$ trigger particles in central $\mbox{Au+Au}$
collisions with emission angle to be independent of the associated
particle $p_{\,\mathrm{T}}$. \cite{STAR-PRL-102-052302-2009} The
observation allows the exclusion of the hypothesis of Cherenkov
gluon emission as the dominant mechanism of production of
away-side peak structure in back-to-back jet configurations at
RHIC. Multihadron correlations show the conversion of energy lost
by associated particles with higher $p_{\,\mathrm{T}}$ into
production of softer hadrons. Direct $\gamma$--hadron correlations
\cite{PHENIX-PRC-80-024908-2009} confirm both conclusions for
path--length dependence of the strength of jet quenching and for
redistribution of missing energy resulted in an increasing of
production of soft particles.

Spectra for identified charged particles exhibits experimental
evidence for different contributions of gluon jets and quark jets
to hadron production in the high-$p_{\,\mathrm{T}}$ domain at top
RHIC energy as well as the absence of the Casimir effect for
parton energy loss in sQGP. \cite{STAR-PRL-108-072302-2012}
Spectrum on $p_{\,\mathrm{T}}$ of direct $\gamma$ in central
$\mbox{Au+Au}$ collisions at top RHIC energy
\cite{PHENIX-PRL-104-132301-2010} shows the inverse slope
$T_{\mathrm{eff}}=221 \pm 19^{\mathrm{stat}} \pm
19^{\mathrm{syst}}$ MeV for $p_{\,\mathrm{T}} > 1$ GeV/$c$, which
can be considered an estimation of the initial temperature of the
final state created in heavy-ion collisions. This temperature is
significantly higher than that predicted by calculations in
lattice QCD for the phase transition from hadronic matter to sQGP
and is in qualitative agreement with hydrodynamical models. The
study of both the spectra and the nuclear modification factor
($R_{\mathrm{AA}}$) for $\pi^{0}$ mesons confirms the strong
dependence of parton energy loss ($\Delta E$) on length of path
($L$) passed by them in hot strongly interacting matter
\cite{PHENIX-PRC-80-054907-2009} and allows the establishment the
relation $\langle \Delta E\rangle \propto \langle L\rangle^{3}$
with help by a hybrid model utilizing pQCD for the hard scattering
and AdS/CFT correspondence for the soft interactions. Spectra
exhibit the significant suppression of heavier mesons with open /
hidden strangeness; \cite{STAR-PLB-673-183-2009} furthermore,
results for non-strange and for strange particles are
complementary in the high-$p_{\,\mathrm{T}}$ domain in some cases.
\cite{PHENIX-PRC-82-011902-2010} Smooth behavior of strangeness
yields, different particle ratios and corresponding
$R_{\mathrm{AA}}$ dependencies on $\sqrt{s_{\mathrm{NN}}}$ at
higher RHIC energies ($\sqrt{s_{\mathrm{NN}}} \geq 62.4$ GeV)
allows the estimation of parameters of the chemical freeze-out.
\cite{STAR-PRC-83-024901-2011} Dilepton mass spectra, in
particular, give the indication in favor of theoretical models
with in-medium broadened $\rho$ contributions in the case of
$\mbox{Au+Au}$ collisions at top RHIC energy.
\cite{STAR-PRL-113-022301-2014} Studying of charged particle
spectra as a function of pseudorapidity for various collision
types and $\sqrt{s_{\mathrm{NN}}}$ results in the following
relation for total multiplicity of secondary charged particles:
$N_{\mathrm{ch}} \propto \ln^{2}s_{\mathrm{NN}}$ for the wide
range $\sqrt{s_{\mathrm{NN}}}=2.7-200$ GeV.
\cite{PHOBOS-PRC-83-024913-2011}

The $R_{\mathrm{AA}}$ of $J/\psi$ with low and intermediate
$p_{\,\mathrm{T}}$ in $\mbox{Cu+Cu}$ and $\mbox{Au+Au}$ is
evidence in favor of models of charmonium production that take
into account the color screening and regeneration and exclude the
model of hydrodynamic flow. \cite{STAR-PRC-90-024906-2014}
Measurements of $J/\psi$ production at
$\sqrt{s_{\mathrm{NN}}}=200$ GeV in the high-$p_{\,\mathrm{T}}$
domain \cite{STAR-PRC-80-041902-2009} show increasing
$R_{\mathrm{AA}}$. In the case of $\mbox{Cu+Cu}$, this observation
contradicts with calculations in the framework of model of
quarkonium dissociation in a strongly coupled liquid using an
AdS/CFT approach, whereas the growth of $R_{\mathrm{AA}}$ for
collisions of mean nuclei are described reasonably by a
two-component model with finite $J/\psi$ formation time. As well
as for light flavors the significant suppression is observed for
yields of $J/\psi$ integrated over transverse momentum in central
$\mbox{Au+Au}$ collisions at the highest RHIC energy.
\cite{STAR-PRC-90-024906-2014} Values of $v_{2}$ for $J/\psi$ at
$p_{\,\mathrm{T}}=2-10$ GeV/$c$ are consistent with zero within
statistical errors. \cite{STAR-PRL-111-052301-2013} Finite $v_{2}$
observed for electrons with $p_{\,\mathrm{T}}=1.3-1.8$ GeV/$c$ and
$p_{\,\mathrm{T}}
> 2.4$ GeV/$c$ from heavy-flavor decays \cite{PHENIX-arXiv-1405.3301}
supposes that heavy quarks interact with the environment at higher
RHIC energies ($\sqrt{s_{\mathrm{NN}}} \geq 62.4$ GeV) but their
degree of thermalization with the medium may depend on
$\sqrt{s_{\mathrm{NN}}}$. Significant suppression is observed for
open charm mesons with intermediate and large $p_{\,\mathrm{T}}$
in $\mbox{Au+Au}$ collisions at the highest RHIC energy,
\cite{STAR-arXiv-1404.6185} which suggests that coalescence can be
considered as mechanisms of hadronization. The first results were
obtained for $J/\psi$ suppression in collisions of asymmetric
beams $\mbox{Cu+Au}$ \cite{PHENIX-arXiv-1404.1873} and for
bottomonium production in various collisions at
$\sqrt{s_{\mathrm{NN}}}=200$ GeV. \cite{STAR-PLB-735-127-2014} The
behavior of $R_{\mathrm{AA}}$ in nuclear collisions exhibits the
complete suppression of excited-state $\Upsilon$ mesons and agrees
with model calculations that include the presence of a sQGP.

Results in the field of correlation femtoscopy show that the
space-time extent of the charged kaon source
\cite{PHENIX-PRL-103-142301-2009} is smaller than that for charged
pions. \cite{STAR-PRC-80-024905-2009} Investigation of
azimuthally-sensitive femtoscopy correlations at top RHIC energy
\cite{PHENIX-PRL-112-222301-2014} indicate that the oscillations
of femtoscopy parameters with respect to the
$3^{\mathrm{rd}}$-order event plane are largely dominated by the
dynamical effects from triangular flow. Estimations of the
space-time extent of the pion emission source in $\mbox{d+Au}$
collisions at $\sqrt{s_{\mathrm{NN}}}=200$ GeV as a function of
kinematic observables show similar patterns in $\mbox{Au+Au}$
collisions and indicate similarities in expansion dynamics in
collisions of various systems at RHIC.
\cite{PHENIX-arXiv-1404.5291} The scaling results for some radii
indicate that hydrodynamic-like collective expansion is driven by
final-state rescattering effects. Study of short- and long-range
multiplicity correlations on pseudorapidity at top RHIC energy
\cite{STAR-PRL-103-172301-2009} established qualitative agreement
with the dual partonic model and pointed to the presence of
multiple parton interactions in $\mbox{Au+Au}$ collisions at
$\sqrt{s_{\mathrm{NN}}}=200$ GeV. Measurements of three-particle
coincidence in pseudorapidity between trigger particle with
$p_{\,\mathrm{T}} > 3$ GeV/$c$ and two lower $p_{\,\mathrm{T}}$
associated particles within small azimuthal separation in
$\mbox{d+Au}$ and $\mbox{Au+Au}$ collisions at top RHIC energy
\cite{STAR-PRL-105-022301-2010} found no correlation between
production of the ridge and production of the jetlike particles,
suggesting the ridge may be formed from the bulk medium itself.
This result, as well as the persistence of the ridge structure
\cite{PHOBOS-PRL-104-062301-2010} over at least $\eta \approx 4$,
indicate that the models attributed the ridge to the medium itself
seem more preferable than the models attributed the ridge to
jet-medium interactions. Accounting for soft particles results in
observation of the ridge in central $\mbox{d+Au}$ collisions at
$\sqrt{s_{\mathrm{NN}}}=200$ GeV also
\cite{PHENIX-PRL-111-212301-2013} that agrees qualitatively with
discovery of similar features of event structure in small system
collisions at the LHC. Thus the measurements at RHIC and LHC are
complementary to each other. System-size dependence of widths of
charge balance functions at $\sqrt{s_{\mathrm{NN}}}=200$ GeV
\cite{STAR-PRC-82-024905-2010} agrees with the hypothesis of
limited diffusion which corresponds with, in turn, the creation of
matter with a very small viscosity, which requires a small mean
free path in $\mbox{Au+Au}$ collisions. Investigations of the
correlation function of transverse momentum
\cite{STAR-PLB-704-467-2011} allow one to confirm the small
viscosity of matter formed in central $\mbox{Au+Au}$ collisions at
top RHIC energy and get the following range of estimations for
ratio of the shear viscosity to entropy density:
$\eta/s=0.06-0.21$. It was also found that transport-based model
calculations show better quantitative agreement with the
measurements compared with models which incorporated only jet-like
correlations. \cite{STAR-PRC-87-064902-2013} On the other hand the
anomalous evolution of two-particle angular correlations on
centrality was observed. \cite{STAR-PRC-86-064902-2012} This
result does not allow one exclude the alternative scenario in
which the heavy-ion collisions in domain, at least, of high RHIC
energies are dominated by minijet structure and the early stage of
space-time evolution of collisions is described by pQCD with
modified fragmentation without such specific properties of created
matter as opacity for few-GeV partons and very small viscosity. It
is important to note that the representation of relativistic
heavy-ion collisions with the creation of matter characterized by
opacity for hard partons and very small viscosity was confirmed by
a wider set of observables than the approach based on minijets. At
present, the former case is the dominant paradigm for the study of
the sQGP.

\section{Fundamental symmetries of QCD at finite temperatures}\label{vao:sec3}
During recent years important progress has been achieved in both
theoretical and experimental studies of fundamental discrete
symmetries of QCD, in particular, $\mathcal{P / CP}$ invariance,
in hot strongly interacting matter. The interplay between the
external Abelian magnetic field with extremely large intensity
(peak $B \sim 10^{15}$ T) and the sQGP created in the heavy-ion
collisions makes local topologically induced violation of
$\mathcal{P/CP}$ invariance in strong interactions possible --
$l\mathcal{TIP}$ effect. One of the possible mechanisms for
manifestation of this effect experimentally is the phenomenon of
electric charge separation along the axis of the applied magnetic
field in the presence of fluctuating topological charge -- the
so-called chiral magnetic effect -- CME. RHIC provides a good
opportunity to study this fundamental physics. Experimentally,
first indication on the CME effect was observed with charged
hadron correlations in high-energy nuclear collisions
\cite{STAR-PRL-103-251601-2009}. Then high statistics study
confirmed the existence of the charge separation effect, and found
that the separation of charge is predominantly orthogonal to the
reaction plane \cite{STAR-PRC-88-064911-2013}, as expected for the
CME. The collision energy dependence of the charge separation
shows the remarkable feature that when the energy is lower than 11
GeV the charge separation approaches zero
\cite{STAR-PRL-113-052302-2014}. This implies the dominance of
hadronic interactions over partonic ones at lower collision
energies and indicates that the onset of such dominance is in the
range $\sqrt{s_{\mathrm{NN}}}=11.5-19.6$ GeV. There is no model
without CME which can describe all experimental results
simultaneously, especially the centrality dependence of same-sign
correlations. Thus most of corresponding experimental results
confirm indirectly the non-trivial topology of QCD vacuum. But
results for multiplicity asymmetry correlations in $\mbox{Au+Au}$
collisions at top RHIC energy show $p_{\,\mathrm{T}}$-dependence
which is unexpected in the framework of CME.
\cite{STAR-PRC-89-044908-2014} These data will stimulate further
developments for background estimations.

\section{Scan of collision energy and future plans}\label{vao:sec4}
Detailed exploration of the phase diagram of strongly interacting
matter and its features is the main goal of the beam energy scan
(BES) program at RHIC. Most results obtained during the first
phase of the program (BES-I) for $\mbox{Au+Au}$ collisions
\cite{STAR-PRL-103-092301-2009} show either a smooth dependence or
absence of the visible changing of corresponding physical
quantities on collision energy. The set of these results can point
on weak changing of contribution of soft processes, at least, in
heavy nuclei collisions at initial energies
$\sqrt{s_{\mathrm{NN}}}=39-200$ GeV. However, it should be
mentioned that at the same time some features observed in
$\mbox{Au+Au}$ collisions,
\cite{STAR-PRL-113-052302-2014,STAR-PRL-112-162301-2014,STAR-PRL-110-142301-2013,STAR-PRL-105-022302-2010}
for example, significant difference in the $v_{2}$ values for
particles and their corresponding antiparticles,
\cite{STAR-PRL-110-142301-2013} allow one to suggest that the
transition occurs at initial energies
$\sqrt{s_{\mathrm{NN}}}=10-20$ GeV in the domain of dominance of
hadronic degrees of freedom over those of quark and gluons.
Precise measurements are planned at top RHIC energy, which will
provide, as expected, new information on the sQGP thermodynamic
and transport properties. For the phase-II of the BES program the
proposed upgrades to RHIC will increase the luminosity for future
low energy runs by a factor of 2 to 20 for
$\sqrt{s_{\mathrm{NN}}}=7.7-20$ GeV, depending on beam energy. The
upgrades to the PHENIX and STAR detector systems will
significantly improve the quality of the measurements. PHENIX has
proposed a new more conventional collider detector, s(uper)PHENIX,
based on a thin-coil superconducting solenoid while, STAR has
proposed two new subsystems, the event plane detector (EPD) and
the inner Time Projection Chamber (iTPC), which will improve the
capabilities of all STAR. Furthermore, STAR plans down to
$\sqrt{s_{\mathrm{NN}}} \approx 3$ GeV and reach to region of
compressed baryonic matter in the framework of the fixed target
part of the phase-II.

\section{Summary}\label{vao:sec5}
In this review some of the main experimental results obtained at
RHIC during the last 5 years have been discussed. The elliptic
flow of charged particles scaled by the initial-state eccentricity
follow a common trend with multiplicity for RHIC and the LHC in
collisions of small systems as well as for large systems. Results
for heavy mesons at top RHIC energy exhibit significant energy
loss by heavy-flavor quarks in hot matter and also indicate
intensive interactions of such quarks with the environment. The
space-time evolution of the ellipticity and triangularity is
observed in angular dependent HBT measurement for $\mbox{Au+Au}$
collisions at the top RHIC energy. Experimental indications of the
local topology-induced parity violation in strong interactions are
observed in nucleus-nucleus collisions at higher RHIC energies.
The deviation from the sQGP signals in the lower energy collisions
indicates the onset of the hadronic degrees of freedom. The
proposed upgrades to RHIC and large experiments (PHENIX, STAR)
will offer the unique opportunity for precise study of the
properties of the sQGP as well as for detailed exploration of the
phase structure of strongly interacting matter. Therefore one can
say that at present the investigation of QCD matter passed from
the ``discovery stage" to ``precision science" studies.

\end{document}